\documentclass[12pt,preprint]{aastex}

\shorttitle{New \ion{He}{2} Quasar Sightlines} 
\shortauthors{}

\begin{document}

\title{A High Yield of New Sightlines for the Study of
Intergalactic Helium: Far-UV-Bright Quasars from
SDSS, {\em GALEX}, and {\em HST}\altaffilmark{1}}

\author{David Syphers\altaffilmark{2},
Scott F. Anderson\altaffilmark{3},
Wei Zheng\altaffilmark{4},
Daryl Haggard\altaffilmark{3},
Avery Meiksin\altaffilmark{5},
Kuenley Chiu\altaffilmark{6},
Craig Hogan\altaffilmark{3},
Donald P. Schneider\altaffilmark{7},
Donald G. York\altaffilmark{8,9}
}

\altaffiltext{1}{Based on observations with the NASA/ESA Hubble Space 
Telescope obtained
at the Space Telescope Science Institute, which is operated by the
Association of Universities for Research in Astronomy,
Incorporated, under NASA contract NAS5-26555.}

\altaffiltext{2}{Physics Department, University of Washington, Seattle, 
WA 
98195; dsyphers@phys.washington.edu}

\altaffiltext{3}{Astronomy Department, University of Washington, Seattle, 
WA 98195; anderson@astro.washington.edu}

\altaffiltext{4}{Department of Physics and Astronomy, Johns Hopkins
University, Baltimore, MD 21218; zheng@pha.jhu.edu}

\altaffiltext{5}{Scottish Universities Physics Alliance (SUPA), Institute 
for Astronomy, University of Edinburgh, Royal Observatory, Edinburgh EH9 
3HJ, United Kingdom}

\altaffiltext{6}{Anglo-Australian Observatory, Epping, NSW 1710, 
Australia}

\altaffiltext{7}{Pennsylvania State University, Department of
        Physics \& Astronomy, 525 Davey Lab, University Park, PA 16802}

\altaffiltext{8}{Department of Astronomy and Astrophysics, The
University of Chicago, 5640 South Ellis Avenue, Chicago, IL 60637}

\altaffiltext{9}{Enrico Fermi Institute,
University of Chicago, 5640 South Ellis Avenue, Chicago, IL 60637}

\begin{abstract}

Investigations of \ion{He}{2} Ly$\alpha$ (304~\AA\ rest) absorption toward 
a half-dozen quasars at $z\sim3-4$ have demonstrated the great potential 
of helium studies of the IGM, but the current critically small sample 
size of clean sightlines for the \ion{He}{2} Gunn-Peterson test limits 
confidence in cosmological inferences, and a larger sample is required. 
Although the unobscured quasar sightlines to high redshift are extremely 
rare, SDSS DR6 provides thousands of 
$z>2.8$ quasars. We have cross-correlated these SDSS quasars with {\it GALEX} 
GR2/GR3 to establish a catalog of 200 higher-confidence ($\sim$70\% 
secure) cases of quasars at $z=2.8-5.1$ potentially having surviving 
far-UV 
(restframe) flux. We also catalog another 112 likely far-UV-bright 
quasars from {\it GALEX} cross-correlation with other (non-SDSS) quasar 
compilations. Reconnaissance UV prism observations with {\it HST} 
of 24 of our SDSS/{\it GALEX} candidates confirm 12 as detected in the 
far-UV, with at least 9 having flux extending to very near the \ion{He}{2} 
break; with refinements our success rate is even higher. Our 
SDSS/{\it GALEX} 
selection approach is thereby confirmed to be an 
order of magnitude more efficient than previous \ion{He}{2} quasar searches, 
more than doubles the number of spectroscopically confirmed clean 
sightlines to high redshift, and provides a resource list of hundreds of 
high-confidence 
sightlines for upcoming \ion{He}{2} and other far-UV studies from {\it HST}. 
Our reconnaissance {\it HST} prism spectra suggest some far-UV 
diversity, 
confirming the need to obtain a large sample of independent 
quasar sightlines across a broad redshift range to assess such issues as 
the epoch(s) of helium reionization, while averaging over 
individual-object 
pathology and/or cosmic variance.

\end{abstract}


\section{Introduction}

A substantial fraction (perhaps most) of the baryons in the Universe did 
not collapse into such dense structures as stars, galaxies, and quasars, 
instead remaining behind in a more dilute intergalactic medium (hereafter, IGM), 
composed primarily of primordial hydrogen and helium 
\citep[e.g., see recent review by][]{mei08}. 
Ultraviolet 
radiation from massive early stars, star-forming galaxies, and quasars 
gradually reionized their surrounding IGM, ultimately ending the ``dark 
ages". Cosmic microwave background studies provide one important indirect 
chronometer for reionization: the {\textit{Wilkinson Microwave Anisotropy 
Probe} ({\textit{WMAP}}) 5-year data \citep{D08} constrain sudden one-time hydrogen 
reionization to $z>6.7$ at $3 \sigma$, and place it at $z=11.0 \pm 1.4$ 
($1 \sigma$). However, the WMAP measures do not strongly distinguish 
between models with one or multiple epochs of reionization; and, although 
they suggest the possibility of an extended reionization epoch, the WMAP 
measures alone do not tightly constrain that possibility \citep{D08}.

Much of the IGM in the nearby Universe is highly ionized, with a hydrogen 
neutral fraction $x_{\mbox{\scriptsize{H I}}}=N_{\mbox{\scriptsize{H I}}}/N_{\mbox{\scriptsize{H II}}} \sim 10^{-5}$ 
\citep{F06}. A sensitive measure of the neutral hydrogen is to examine 
quasar spectra for the saturated trough shortward of Ly$\alpha$ caused by 
even a modest amount of neutral IGM gas, i.e., the classic Gunn-Peterson 
effect \citep{GP65}. Somewhat in contrast to initial hints from the first 
few $z\sim$6 quasars examined, data from $\sim$20 now-studied 
high-redshift QSO sightlines, primarily from objects discovered in the Sloan Digital Sky Survey 
\citep[SDSS;][]{Y00}, suggest that the reionization of \ion{H}{1} 
likely happened before $z\sim6$ \citep{F06}.  The \ion{H}{1} Gunn-Peterson trough 
is evident in some $z\sim6.3$ quasars \citep{B01,F06}, suggesting at 
least that the ionization of hydrogen is patchy and may be rapidly 
evolving in the $z>6$ regime. However, there is marked sightline variance 
in the ensemble of quasar \ion{H}{1} Gunn-Peterson measurements and their 
interpretations \citep[e.g.][]{W03,F06,T06,bol07,bec07}. Moreover, some 
high-redshift galaxy studies also tend to favor higher redshifts for \ion{H}{1} 
reionization \citep[e.g.][]{hu02,rho04}, as galaxies with strong (and 
inferentially, largely unabsorbed) \ion{H}{1} Ly$\alpha$ emission are seen out to 
(at least) $z\sim7$ \citep[e.g.][]{iye06}.

Although hydrogen dominates both the number and mass fraction of baryons, 
helium is the most abundant absorber in much of the IGM accessible to 
current study. The \ion{He}{2} reionization epoch is likely delayed 
versus that of \ion{H}{1} because hard-ionizing sources such as AGN had 
to form to produce the higher energy photons needed \citep[e.g.][]{W03b}. 
Even under the hard photoionizing conditions at $z\sim 2-4$, \ion{He}{2} 
outnumbers \ion{H}{1} by a factor of $\eta \sim 50$-100 \citep{fec06}, and 
has a higher opacity by $\eta/4 > 10$. The \ion{He}{2} Ly$\alpha$ 
transition is 
thus the more sensitive, higher opacity, empirical probe of much of the 
highly ionized IGM---as compared to \ion{H}{1} which over most of sampled 
redshift space (between the \ion{H}{1} Ly$\alpha$ forest lines) is extremely 
sparse. \ion{He}{2} studies allow cosmological inferences about the epoch(s) of 
reionization, the intensity and spectrum of the ionizing background 
radiation, and measures of the cosmic baryon density in the IGM. However, 
\ion{He}{2} Ly$\alpha$ absorption occurs in the far UV (304\AA\ restframe), only 
observable from space in a rare fraction of high redshift quasars which 
lie by chance along clean lines of sight with little foreground 
absorption.

Jakobsen et al.'s pioneering work \citep{J94} with the {\it Hubble Space Telescope}
(hereafter, {\it HST}) provided the 
first detection of a \ion{He}{2} Gunn-Peterson trough, in Q0302-003 
($z=3.29$).  Yet even following a further decade of effort, only four 
quasar sightlines suitable for \ion{He}{2} Gunn-Peterson studies had been 
confirmed---though each was subjected to intensive UV spectral campaigns 
from {\it HST}, {\it HUT}, and/or {\it FUSE}. In redshift order these 
objects are: HS~1700+64 at $z=2.72$ \citep{D96,fec06}, HE~2347-4342 at $z=2.88$ 
\citep{R97,kri01,sme02,Z04a,S04}, PKS~1935-692 at $z=3.18$ \citep{A99}, and 
Q0302-003 at $z=3.29$ \citep{J94,H97,H00}. Broadly, the various 
intensive UV spectroscopic studies of these four quasars indicate that the IGM 
\ion{He}{2} optical depth increases markedly from $\tau\sim 1$ near $z=2.5$ to 
$\tau >4$ at $z=3.3$, and are also consistent with theoretical notions 
\citep{haa96}, and other indirect evidence \citep{son96,the02}, that 
helium reionization likely occurred near $z\sim$3 (delayed vs. that of 
hydrogen at $z>$6). These intensive follow-up studies have also provided 
a handful of empirical measures of the flux in the ionizing background 
radiation at $z\sim$3 and the baryon density ($\Omega_g\sim0.01$) in the 
IGM \citep[e.g.][]{H97,Z04a,T07}. A predicted signature of the \ion{He}{2} 
reionization epoch is a damping absorption profile {\it redward} of \ion{He}{2} 
\citep{mir98,mad00}; this feature was unseen in these four intensively 
studied cases at $z<3.3$, but perhaps merely because they are at too low 
redshift (when \ion{He}{2} ionization was well underway).

From later {\it HST} ultraviolet follow-up of selected optically bright 
high-redshift quasars, \citet{rei05} identified QSO 1157+3143 (at 
$z=3.0$) as another \ion{He}{2} quasar, and our group added three more clean 
quasar sightlines that 
extend helium IGM studies to the highest redshifts heretofore sampled 
\citep{Z04b}. From an {\it HST} SNAP survey we first confirmed SDSS 
2346-0016 ($z=3.5$), SDSS 1711+6052 ($z=3.8$), and SDSS 1614+4859 
($z=3.8$) as excellent new clean sightlines for helium IGM studies; our 
6\% selection efficiency, though modest, was among the highest then obtained.  
Moreover, our lengthy follow-up with the Advanced Camera for 
Surveys, Solar Blind Channel (ACS SBC) prism (taken after the failure of the Space Telescope Imaging Spectrograph [STIS])
for two of these, SDSS 2346-0016 and SDSS 1711+6052, shows an 
especially intriguing result: the $z=3.8$ case shows a surprisingly
prominent absorption profile redward of the \ion{He}{2} break location 
\citep{Z08} that is not seen in the $z=3.5$ case. The profile in 
SDSS 1711+6052 is stronger and wider than commonly expected for the damped 
profile signature of the IGM, so in that paper we also consider 
additional explanations including absorption in a high-redshift intense 
star-forming region. Whatever the explanation, the strength of this 
surprising feature demands further scrutiny in future high-quality UV 
spectra 
toward SDSS 1711+6052, as well as other comparably high-redshift \ion{He}{2} 
quasar sightlines.

Motivated by the previous small sample size that limits confident global 
conclusions, and the potentially intriguing results at 
higher redshifts \citep[e.g.,][]{Z08},
here we detail our successful efforts to markedly 
increase the number of new quasar sightlines suitable for the study of 
\ion{He}{2}, across a broad redshift range.
Starting from the very large sample of 7800 SDSS quasars at suitable 
redshift, we use the {\textit{Galaxy Evolution Explorer}} ({\it GALEX}) archive to then cull to a new catalog of 200 
higher-confidence far-UV detections of quasars at $z>2.78$. An additional 
112 quasar sightlines likely clean to high redshift are also similarly 
identified from cross-correlation of {\it GALEX} with other (non-SDSS) quasars. 
Our catalogs of likely far-UV-bright (restframe) quasars, presented in section 2,
provide appropriate lists from which to draw for follow-up observations.

In section 3 we present reconnaissance UV observations with the 
{\it HST} ACS prism of 24 SDSS/{\it GALEX} candidate far-UV-bright quasars. 
Among these, we confirm 12 as bright in the far UV, with at least 9 
evidently having flux all the way down to (very near) the \ion{He}{2} break.

As we discuss in section 4, our SDSS/{\it GALEX} selection approach is thus 
found to be an order of magnitude more efficient than previous \ion{He}{2} 
quasar searches, and herein we more 
than double the number of clean sightlines spectroscopically confirmed as 
suitable for helium IGM studies. Moreover, our reconnaissance {\it HST} 
UV prism spectra suggest diversity, further confirming the need to obtain 
a large sample of independent quasar sightlines over a broad redshift 
range to understand cosmological variance and/or possible 
individual-object peculiarity.

\section{SDSS/{\em GALEX} Selection and Catalogs of Far-UV-Bright 
Quasars}

In order to identify a large new sample of clean sightlines to 
high redshift, we cross-correlate the very large SDSS 
quasar sample with the 
{\textit{GALEX}} catalog of UV sources. SDSS is fundamental to the 
process, with its optical multicolor photometry and multiobject 
spectroscopy efficiently identifying a very large number of quasars out to 
high redshifts \citep{ric02}. The broadband {\textit{GALEX}} catalog observations 
efficiently confirm which quasars are likely to have flux well into the 
far UV (restframe).

\subsection{SDSS Quasar and {\em GALEX} Input Catalogs}

The SDSS provides an optical digital imaging and spectroscopic data bank 
of a region approaching $\sim10^4$~deg$^2$ of sky, primarily centered on 
the north Galactic polar cap. Imaging and spectroscopic data are obtained 
by a special purpose 2.5m telescope, at Apache Point Observatory, New 
Mexico, equipped with a large-format mosaic camera that can image of 
order $10^2$~deg$^2$ per night in 5 filters ($u,g,r,i,z$), along with a 
multifiber spectrograph that obtains 640 spectra, simultaneously, within 
a 7~deg$^2$ field.  The imaging database is used to select objects for 
the SDSS spectroscopic survey, which includes 
spectrophotometry ($\lambda/\Delta\lambda\sim1800$) covering 
3800-9200\AA\ for $10^6$ galaxies, $10^5$ quasars, and $10^5$ stars. 
Technical details on SDSS hardware and software, and astrometric, 
photometric, and spectral data may be found in a variety of papers: e.g., 
\citet{fuk96}, \citet{gun98}, \citet{lup99}, \citet{Y00}, \citet{hog01}, 
\citet{sto02}, \citet{smi02}, \citet{P03}, \citet{ive04}, and 
\citet{gun06}. A description of the most recent SDSS public data release 
(Data Release 6; hereafter, DR6) is given by \citet{ade08}.

The 5 SDSS optical filters plus multiobject spectra efficiently select 
quasars out to high redshifts. Here we 
focus on 
quasars from the SDSS 
catalog with redshift $z>2.78$, for which the \ion{He}{2} Ly$\alpha$ line 
at 304\AA\ would then be observed at $>$1150\AA. Observations in this 
wavelength regime must, of course, take place in space, and are thus 
currently limited mainly to {\textit{HST}}, although some fundamental 
observations 
of the brightest objects have been possible with FUSE 
\citep[e.g.][]{kri01,S04,Z04a}, and 
{\it HUT} \citep{D96} as well. Our specific choice of $z>2.78$ 
here is motivated, in part, by the upcoming {\it HST} Service Mission 4; a 
refurbished STIS might be anticipated to be useful down to $\sim$1150\AA, 
and the Cosmic Origins Spectrograph (COS) may even still have rather good 
response at this wavelength.

In the 5700~deg$^2$ of sky with spectroscopic coverage in SDSS Data 
Release 5 (DR5), there are $>$77,000 spectroscopically confirmed quasars 
\citep{S07}, among which are nearly 6600 at $z>2.78$ (e.g., see Figure 1). 
We include all such high-redshift confirmed DR5 quasars in our 
cross-correlation with {\it GALEX}. In addition, there are another 2300 SDSS 
candidate quasars potentially in this redshift regime in the further 
spectroscopic coverage of SDSS DR6. There is no officially vetted DR6 
quasar catalog, so our search for the latter used the SDSS DR6 CasJobs 
table {\texttt{QSOConcordanceAll}}, which contains most of the likely 
quasar candidates. Given the likely contamination (usually by real quasars 
at incorrect redshifts) of this latter, not-yet-vetted set, we examined
these additional DR6 spectra to exclude non-quasars and verify redshifts
before matching to the {\it GALEX} catalogs.
Our starting SDSS catalog thereby includes about 7800 spectroscopically 
confirmed SDSS quasars at $z>2.78$.

However, identifying the small fraction of rare sightlines 
suitable in practice for \ion{He}{2} studies demands UV observations as well. In 
order to be useful in constraining \ion{He}{2} reionization and 
absorbers, the observed quasar spectra must have detectable flux down to the \ion{He}{2} Ly$\alpha$ 
break, $304$\AA. Unfortunately, the UV fluxes of high-redshift objects 
are attenuated by accumulated absorption in numerous intervening 
HI Ly$\alpha$ lines \citep[the Lyman Valley;][]{mol90}. Also, \ion{H}{1} 
absorption from intervening Lyman-limit systems often severely cuts flux 
below 912\AA$(1+z_{abs})$. Only a few percent of random quasar sightlines to 
such high redshift are anticipated to be clear of this UV-obscuring 
neutral hydrogen \citep[e.g.,][]{mol90,rei05}.

Hence, we cull through the 7800 SDSS $z>2.78$ optical quasars to find 
those most likely to be clean sightlines to high redshift, by 
cross-correlation with the {\it GALEX} catalogs. {\it GALEX} is 
performing a large-scale UV broadband imaging survey \citep{mar05} in both 
the FUV ($\sim$1350-1750\AA) and NUV ($\sim$1750-2800\AA) bands. There are 
three surveys: the all-sky survey (AIS) extends to 
$m_{\mbox{\scriptsize{AB}}}\sim$21, and the much smaller area medium and 
deep surveys (MIS and DIS) extend to $m_{\mbox{\scriptsize{AB}}}\sim$23 and 
$m_{\mbox{\scriptsize{AB}}}\sim$25, respectively. The {\it GALEX} GR2 
and GR3 catalogs (henceforth GR2/GR3) are complementary data sets 
processed 
with the same pipeline and constituting one catalog \citep{M07}. GR2/GR3 
covers $\sim$13,500 deg$^2$ of sky, but
the {\it GALEX} sky coverage does not contain all the SDSS 
DR5/DR6 spectroscopic sky coverage. Among the 7800 SDSS $z>2.78$ quasars 
discussed above, {\it GALEX} GR2/GR3 provides UV imaging coverage for 4704 
(3913 from the vetted DR5 catalog, and 791 more from DR6 only). It is 
these 4704 SDSS quasars that are actually matched to {\it GALEX} GR2/GR3 
UV 
catalogs, and considered further in the next subsection. Although the 
broadband {\it GALEX} observations do not provide sufficient information to 
conclusively confirm far-UV flux all the way down to \ion{He}{2}, they at 
least may confirm which of our quasars have flux well into the restframe 
far UV, i.e., to UV wavelengths where most random quasar sightlines are 
obscured by intervening clouds.

\subsection{SDSS/{\em GALEX} Cross-Correlation}

A key parameter in the cross-correlation of catalogs is the maximum 
SDSS/{\it GALEX} positional match radius. Of course this depends on the 
FWHM of FUV and NUV {\it GALEX} point-spread functions, as noted below. But 
this also depends in part on the astrometric accuracy of the surveys. 
Since the SDSS astrometric accuracy is better than $0.1''$ \citep{P03}, 
the modest astrometric limiting factor is {\it GALEX}, with accuracy of 
about $0.5''$ rms \citep{M07}.
(A match radius based on only this formal error would be an underestimate, 
since, unsurprisingly, the non-gaussian separation distribution has a long tail.)
The optimal match radius also depends on the surface density of 
the objects of interest; we are somewhat inclusive in our choice because 
there are comparatively few \ion{He}{2} candidates.
This is weighed against the background rate of {\it GALEX} sources contributing as false matches. 
Because our targets are found on a unique mixture of AIS, MIS, and DIS 
images, we preferred a direct measure of the best match radius for our 
particular application. To characterize the background rate of 
happenstance positional coincidences, we performed matching between mock 
SDSS quasar catalogs and the {\it GALEX} catalog, by randomly shifting 
from our actual SDSS quasar positions. These shifts were in various 
mixtures of RA and Dec, and of a few arcminutes each. This is large 
enough to be well outside the original match radius, but generally to 
remain on the same {\it GALEX} $1.25\degr$-field-of-view tile image as 
the true search (and hence sampling to a very similar {\it GALEX} depth).

We initially considered a match radius extending out to $r<6''$ (in both 
actual, and in mock Monte Carlo, catalog matches); this starting radius 
was 
chosen because the GR2/GR3 spatial resolution is about $5.3''$ FWHM in 
the NUV 
\citep{M07}. We verified that the spurious matches in the Monte Carlo 
shifts were uniform in $r^2$, as should be the case. This uniform 
background is shown in Figure 2
overplotted on our actual cross-correlation of SDSS/{\it GALEX} catalogs. 
From our mock-catalog Monte Carlo shifting considerations, we determined 
that a cut at about $r<3''$ match radius might be effective in avoiding 
most spurious matches; at match radius $r<3''$, our Monte Carlos suggest 
that about $70 \pm 2 \%$ of cases are likely real SDSS quasar/{\it GALEX} 
UV source associations, with 30\% spurious matches. Very few real associations are outside of this 
match radius, although UV sources that are too faint to be extracted into 
the {\it GALEX} catalog will of course go unnoticed. (For comparison, we 
estimate that $\sim$20\% of true matches found within $3''$ would be outside of a $2''$ 
radius.)

\citet{M07}, in their review of {\it GALEX} performance, recommend 
cross-correlating {\it GALEX} and catalogs of higher astrometric 
accuracy with a maximum search radius of $2''$, for many purposes. We 
choose the slightly more generous $r<3''$ limit for several reasons. 
First, \citet{M07} considered mainly UV-bright sources, while we are 
concerned with many sources near the UV detection limits; our own 
analysis and that of others \citep[including][]{M07} reveal, unsurprisingly, that 
higher S/N detections tend to have higher astrometric accuracy. Second, 
we have a number of quasars detected in FUV only, and the FUV astrometric 
precision is worse than that of the NUV \citep{M07}. Third, our comparison 
with {\it HST} reconnaissance observations of some of the objects on our 
list (see section 3)
ultimately empirically confirms that $3''$ is an appropriate balance 
between getting most true matches and excluding most spurious ones. 
Fourth, a number of other SDSS/{\it GALEX} studies \citep[e.g.,][]{agu05,tra07} 
also prefer allowing offsets out to about $3''$.

Among the 4704 SDSS DR5/DR6 $z>2.78$ quasars within the {\it GALEX} 
GR2/GR3 
area, we find 419 unique matches within a match radius of $r<6''$. (We preferred closer matches in 
the cases where more than one UV source was within the search radius of 
the SDSS optical object.) Among 
these 419, there are
212 within our ultimately preferred $r<3''$ match radius.

Finally, we inspected both the SDSS and {\it GALEX} images for the 
positionally preferred 212 cases, for any 
obvious problems. No further objects were removed based on SDSS image 
concerns, but several cases were flagged as having very near optical 
neighbors. These projected neighbors may lead to complications in at least 
two ways for helium IGM studies: first, if the projected neighbor is an 
intervening galaxy, then the quasar UV spectroscopic sightline might be 
impacted by low-redshift \ion{H}{1} or other absorption from the galaxy (or 
group/cluster in which the projected neighbor galaxy resides); second, the 
neighboring optical object might actually be the {\it GALEX} source, 
rather than 
the quasar. We also inspected the {\it GALEX} images, both NUV and FUV, 
looking for catalog matches where the UV object appeared to be an artifact. 
(Note the GR2/GR3 catalogs have artifact flags for each object which we 
consulted as well.) After removing {\it GALEX} artifact matches, and those too 
close to {\it GALEX} tile edges or other artifacts to be certain of the 
identification, we were left with the 200 objects that constitute our final 
catalog of higher confidence SDSS/{\it GALEX} quasars at $z>2.78$.

All of the previously confirmed \ion{He}{2} quasars in our SDSS position 
and redshift
footprints are recovered in our higher confidence catalog in a blind 
fashion. These include the quasars: Q0302-003
\citep[$z=3.29$;][]{J94}, SDSS J2346-0016 \citep[$z=3.51$;][]{Z04b}, SDSS
J1614+4859 \citep[$z=3.80$;][]{Z05}, and SDSS J1711+6052
\citep[$z=3.83$;][]{Z08}; see magenta stars in Figure 1. In addition, 
OQ172
\citep[$z=3.54$;][]{lyo92,L94,L95,jak93}, a very bright quasar 
considered and
rejected as a \ion{He}{2} candidate in very early {\it HST} UV studies (with a 
break 
just redward of \ion{He}{2}
Ly$\alpha$) is also recovered in our catalog (see red star symbol 
at $z=3.54$ in 
Figure 1), further reassuring that we 
are 
not missing many potentially interesting objects.

\subsection{Catalog of Far-UV-Bright SDSS/{\em GALEX} Quasars}
\label{sec:targetdata}

The 200 SDSS quasars at $z>2.78$, associated at higher 
($\sim$70\%) confidence with {\it GALEX} UV sources (3$''$ match radius) are cataloged 
in Table \ref{tab:TargetList}. Selected basic data are also displayed in 
Figures 1 and 3 (see medium, filled black circles depicting the 200 
quasars).

In Table \ref{tab:TargetList}, the catalog columns for these 
candidate far-UV-bright 
(restframe) quasars are as follows.
{\textit{Columns 1-5}}---Names, positions (RA, dec; J2000), redshifts, 
and magnitudes are all from SDSS.
The redshift is from the automated pipeline, but we verified 
it to be reasonably accurate in each case, via our by-eye examinations of 
the spectra.
{\textit{Columns 6-7}}---The {\it GALEX} catalog provides fluxes in $\mu$Jy, which we have converted to 
erg~s$^{-1}$~cm$^{-2}$~\AA$^{-1}$ using the effective wavelengths of the {\it GALEX} broadband filters: 2267\AA\ for NUV, and 1516\AA\ for FUV.
{\textit{Column 8}}---Inspection flag. This is a 4-bit integer flag, 
reflecting our by-eye inspection SDSS/{\it GALEX} images and SDSS spectra. We 
inspected the SDSS images, and if there was a galaxy or possible galaxy 
nearby in projection (typically within $10''$), or a very near blue object 
(typically within $5''$), this flag gets a value of 1. A value of 2 
indicates that we saw a probable Lyman-limit system (LLS) or damped 
hydrogen Ly$\alpha$ absorber (DLA) in the optical SDSS spectrum. A value 
of 4 is 
given to those objects that are broad absorption line quasars 
(BALQSOs; defined herein as having broad \ion{C}{4} absorption in excess 
of 
2000~km~s$^{-1}$), 
and a value of 8 is given to those objects that are borderline on 
our BALQSO criterion. These flags are additive, and hence a flag of 6 
here would mean a BALQSO that also has a LLS or DLA.
{\textit{Column 9}}---{\it HST} observation flag. 
The observation flag is 0 if the object has not been observed with {\it HST}, 1 if it has been observed 
with 
FUV (to near 304\AA\ rest) spectroscopy through {\it HST} cycle 16, and 2 if it has been 
observed 
less conclusively, i.e., NUV spectroscopy, UV imaging, or pre-COSTAR observation.

\subsection{Catalog of Other Far-UV-Bright {\em GALEX} Quasars}

In addition to the SDSS quasars, we similarly matched the quasar catalog 
of \citet{ver06} to {\it GALEX} GR2/GR3 as well. There are 1144 $z>2.78$ 
quasars in \citet{ver06} that are not in SDSS DR6, of which 705 are in the 
{\it GALEX} footprint. We performed the same match radius determination 
described above on this non-SDSS quasar compilation, and found a match 
radius of $3.5''$ was a better compromise in this case, perhaps reflecting 
the more heterogeneous astrometric heritage of these quasars. Within this 
radius, there were 112 matches ($84 \pm 1$\% likely true matches) to {\it GALEX} GR2/GR3 sources; hereafter, 
we refer to these as the VCV/{\it GALEX} quasars.
(One additional match was discarded since it also matched a low-redshift SDSS quasar; the SDSS redshift was verified to be correct.)
Although accurate optical magnitudes on a common system are not uniformly 
available for these VCV/{\it GALEX} quasars, they may be
slightly brighter in the optical than the SDSS sample, and this may 
contribute 
some to the higher match rate to GALEX. However, given the steep 
increase in intervening obscuration with redshift, it seems more 
plausible that the 
higher match rate is
mainly a redshift effect. As may be seen in Figure~4,
the redshift distribution of the VCV/{\it GALEX} quasars favors
systematically lower-redshifts than the SDSS quasar sample, which 
generally 
extends to quite high-redshift \citep[e.g.,][]{and01};
for example, only about 1/4 of the VCV/{\it GALEX} matches are at 
$z>3.05$ compared to about 1/2 of the SDSS/{\it GALEX} matches.
No inspection flag was readily 
feasible for these additional quasars, given 
the difficulty of obtaining uniform optical images and spectra for this 
heterogeneous sample. Our catalog of VCV/{\it GALEX} far-UV-bright 
(restframe) quasars is provided in Table 
\ref{tab:TargetListVeron}.

In Table \ref{tab:TargetListVeron}, the catalog columns for these 
far-UV-bright (restframe) quasar candidates are as follows. 
{\textit{Columns 1-5}}---Names, positions (RA, dec; J2000), redshifts, and 
magnitudes are all from \citet{ver06} and the specific references cited 
therein. The optical magnitude provided is a V magnitude unless otherwise 
marked as: photographic (*), blue (B), red (R), infrared (I J K), 
or photographic O- or J- plates (O). {\textit{Columns 6-7}}---The {\it 
GALEX} catalog provides fluxes in $\mu$Jy, which we have converted to 
erg~s$^{-1}$~cm$^{-2}$~\AA$^{-1}$ using the effective wavelengths of the 
{\it GALEX} broadband filters: 2267\AA\ for NUV, and 1516\AA\ for FUV. 
{\textit{Column 8}}---{\it HST} observation flag.
The observation flag is 0 if the object has not been observed with 
{\it HST}, 1 if it has been observed with FUV (to near 304\AA\ rest) spectroscopy through 
{\it HST} cycle 
16, and 2 if it has been observed less conclusively, i.e., NUV spectroscopy, UV imaging, or pre-COSTAR observation.

Again, reassuringly, we recover in a blind fashion
both additional previously confirmed \ion{He}{2} quasars 
that might be anticipated to be among our VCV/{\it GALEX} 
finds, PKS~1935-692 ($z=3.18$) and HE~2347-4342 ($z=2.88$)
(see magenta stars in Figure 3). In addition,
UM670 \citep[$z=3.16$;][]{jak93,L94}, another very bright quasar
considered and
rejected as a \ion{He}{2} candidate in very early {\it HST} UV studies (with a
break
just redward of \ion{He}{2}
Ly$\alpha$) is also recovered in our catalog (see red star symbol in
Figure 3).
Finally, note that there are only two further 
venerable \ion{He}{2} quasars (see large green circles in Figure 3, lacking a 
central black point) to
account for that do not appear in our catalogs. But these two are
absent from our catalogs for fully anticipated
reasons: HS~1157+3143 \citep{rei05} is not within the {\it GALEX} GR2/GR3 footprint, 
while HS~1700+64 \citep{D96} falls just below 
our catalog redshift cut.

The 200 SDSS and 112 other $z>2.78$ quasars cataloged in this and the 
previous subsections represent significant 
resource lists for far-UV (restframe) studies of quasars. Our 
UV-bright quasars may find applications to a range of studies from
the study of far-UV SEDs, emission lines, and BALs, to study of 
low-redshift hydrogen absorbers, to our interest in \ion{He}{2}.
In the next 
section we 
focus on preliminary reconnaissance follow-up observations from {\it HST}
of selected SDSS/{\it GALEX} quasars, 
relevant to our primary aim of identifying a large new sample of 
independent, clean 
sightlines potentially suitable for helium IGM studies.

\section{{\em HST}/ACS-Prism Confirmation of a High Yield of Far-UV-Bright 
Quasars}
\label{sec:ACSRecon}

\subsection{Reconnaissance {\em HST} UV Spectroscopy}

In {\it HST} cycles 15 and 16 (GO programs 10907 and 11215), we 
conducted programs of brief reconnaissance UV spectra of some of our 
SDSS/{\it GALEX}  
quasars; we have observed 24 of our SDSS/{\it GALEX} candidate 
quasars with brief UV ACS/SBC prism spectra to verify 
end-to-end their potential suitability for future detailed \ion{He}{2} 
studies. Our {\it HST} target list was originally based 
on the then-available {\it GALEX} GR1 catalogs, supplemented later by 
expansions to {\it GALEX} GR2/GR3 (as well as continuing expansions in the 
SDSS quasar catalogs). Three of the 24 {\it HST} targets 
appeared as UV sources in the {\it GALEX} GR1 catalog, but 
not (at least within a 6$''$ offset) in the later GR2/GR3 catalog. However, 
the bulk we observed with {\it HST} are indeed in {\it GALEX} GR2/GR3 catalogs, 
also providing 
an end-to-end test of our SDSS/{\it GALEX} selection approach discussed in 
section~2. About half of our {\it HST} targets had
SDSS/{\it GALEX} positional offsets (at least in the GR2/GR3 catalog) in the 
range $r<3''$,
and about half beyond $r>3''$. 

The reconnaissance UV spectra (plus accompanying direct UV images 
described below) encompass 2 {\it HST} orbits each, with a 
total of 6 individual prism exposures for each object; 
individual exposures are about 600-800s each, with typical total exposure 
times of about 3-5~ksec for the coadded spectrum for each quasar. 
The prism PR130L
was used to limit complications (e.g., markedly enhanced backgrounds)
due to geocoronal Ly$\alpha$, and effectively covers the wavelength
range of about 1250-1750\AA. (In some cases we cautiously use data 10
or 20\AA\ blueward, to help constrain the behavior of the lower-redshift
objects. Although the sensitivity of PR130L falls
quickly below 1250\AA, it is not zero.)
Our 
{\it HST} observations and reductions also involve, for each object, four $\sim$100s 
direct UV image exposures with ACS/SBC using the F150LP filter, needed 
to establish the prism spectra wavelength zero points.

A total of 12 of the targeted 24 quasars are verified with {\it HST} to 
have 
detectable far-UV (restframe) flux in the ACS PR130L wavelength regime. 
While not all of those are suitable for \ion{He}{2} IGM studies, the 
majority are, and our {\it HST }reconnaissance results confirm 
the very high yield of far-UV-bright quasars obtainable from 
SDSS/{\it GALEX} cross-correlations for $r<3''$ offsets. Only two of the 
$r<3''$ cases observed in our {\it HST} reconnaissance 
lack a UV flux detection in our prism data. Moreover, one of 
those two is
actually also excluded from our current catalog of 200 higher confidence far-UV 
bright quasars discussed in section~2; it was among the $r<3''$ positional 
matches, but excluded in section 2 at the final culling stage when 
examining {\it 
GALEX} images, as a possible {\it GALEX} artifact. (This exclusion was based on GR2/GR3 data not available at the time of {\it HST} observation.)

The 12 
far-UV-bright confirmations from our {\it HST} prism reconnaissance are 
listed in 
Table \ref{tab:ACSObs}.
Tabulated are the following columns.
{\textit{Columns 1-4}}---Names, positions (RA, dec; J2000), and redshifts are all from 
SDSS. The redshifts are from the automated pipeline, but we verified them 
to be reasonably accurate in our by-eye examination (only one case, 0808+4550,
required a redshift adjustment, by $0.05$).
{\textit{Columns 5-6}}---The {\textit{GALEX}} catalog provides fluxes 
in $\mu$Jy, which we have converted to 
erg~s$^{-1}$~cm$^{-2}$~\AA$^{-1}$ using the effective wavelengths of 
the {\textit{GALEX}} broadband filters: 2267\AA\ for NUV, and 1516\AA\ 
for FUV. The errors quoted are those given by the {\textit{GALEX}} 
catalog.
{\textit{Columns 7-8}}---Exposure times and date the target was 
observed with HST ACS. The exposure times indicate how much data was used 
to produce each final coadded spectrum.
In cases where we discarded some particularly noisy or misaligned 
exposures, this differs a little from the actual total exposure time.
{\textit{Column 9}}---Indicates whether or not there is a sharp break 
at a wavelength consistent with that expected for \ion{He}{2} 
Ly$\alpha$. Quasars identified in the table as likely having a break at 
\ion{He}{2}, conservatively must show a break of comparable width to (or a 
little wider than, as expected for a proximity feature) the 
prism instrumental resolution.
(Other objects listed as uncertain in Table~2 may 
nonetheless also still be of potential interest for future \ion{He}{2} 
studies.) Note that no objects have 
obvious significant flux that extends blueward of the expected break 
location, although this is not well constrained in the lower-redshift cases.

\subsection{Basic {\em HST} UV Prism Spectral Reductions}

For the 12 cases with ACS/{\it HST} prism detections, we reduced the ACS 2-D 
spectra with the aXe software, version $1.6$ 
\citep{K06}. We consulted both SExtractor and IRAF tasks applied to the 
direct 
F150LP images to determine the positional information input needed for 
aXe.
The spectrum from each individual 600s-800s exposure was 
extracted separately, 
with local background subtraction. In almost every case, our targets were 
the 
only objects evident in the UV {\it HST} exposures.
For each quasar we coadded the multiple individual 600s-800s 
extracted spectra. This was complicated by the lack of a common 
wavelength scale, which we accounted for by fitting each spectrum with 
a high-order cubic spline (with knots determined by the first 
exposure's wavelength scale).
These fits matched the original data to $\lesssim 0.1$-1\%, and thus
were a negligible source of error.
We examined each 
individual exposure's spectrum, and discarded those that were vastly 
more noisy than the average.
These fitted individual exposure spectra, now with a common wavelength 
scale, were then averaged for each quasar.
The discarded exposures were about $\sim 10$\% of the total 
observation time, and no more than $\sim 25$\% for any given quasar.

Relative fluxes in these initial prism reductions should be reasonably 
good, but possible systematic errors in absolute spectrophotometry 
zero-points are 
a residual concern and under continuing study (though they do not impact any 
conclusions herein). For example,  
errors in determining the target's position in the F150LP images would
lead to systematically lower fluxes since the extraction trace would 
not be exactly along the spectrum.
Another factor that affects flux measurements is the weighting 
used during the extraction from the 2-D spectrum. Here, we have chosen to 
use
optimal weighting \citep{H86} as implemented by aXe, which uses lower
weights for pixels farther from the spectrum's trace (on the axis
perpendicular to dispersion) to improve S/N; such optimal weighting may 
tend to also underestimate the flux zeropoints (though alternate 
weightings at least give very similar {\it relative} fluxes).
Finally, there appears to be some 
possible systematic underestimate of the spectral flux compared to 
the {\textit{GALEX}} broadband data.

The extracted, coadded spectra of the 12 far-UV-bright quasars 
confirmed from {\it HST} are presented in Figure 5a and 5b.
The vertical dotted line shows 
the 
wavelength of expected IGM \ion{He}{2} absorption for the given quasar 
redshift. The absolute flux calibration is uncertain, as noted above, but 
the relative flux calibration is more secure. The width of the bins 
is 
about half of the instrumental resolution at the given wavelength (for 
PR130L, $R$ ranges from about 350 at 1250\AA\ to about 60 at 1670\AA).

\section{Discussion: A High Yield of New Sightlines for Helium Studies}
\label{sec:ACSxcor}

The ACS prism data shown in Figure 5 confirm that our catalogs presented 
in 
section 2
are successful in including most real matches and 
excluding most false matches. 
Among the 24 SDSS/{\it GALEX} quasars with reconnaissance {\it HST} spectra, about 
half have SDSS/{\it GALEX} positional offsets in the range $r>3''$, but 
none of those were confirmed as having significant far-UV (restframe) flux in 
the ACS prism data; this further confirms our choice to catalog only the
higher confidence cases with $r<3''$ in section 2.
Among 
our 200 higher-confidence SDSS cases cataloged, we have observed 11
with ACS/prisms.
Ten of the 11
such higher-confidence cases from our cycle 15-16 observations are 
confirmed as far-UV-bright (restframe) in 
our reconnaissance ACS spectra. In fact, the {\it HST} reconnaissance ACS prism 
observations were first initiated based on {\it GALEX} DR1 catalogs, and we 
also have two additional quasars (SDSS 1009+3917 and 
SDSS 1137+6237) 
that appear as {\it GALEX} sources at 
$r<3''$ in {\it GALEX} DR1, but not in GR2/GR3; both of these quasars are also 
confirmed in
our {\it HST}/prism spectra to be bright in the far UV.

Of the total of 12 cases definitively verified as far-UV-bright 
from {\it HST} prism spectra, at least 9 show surviving flux all the way down 
to very near 
the wavelength expected for IGM \ion{He}{2} Ly$\alpha$, and are 
potentially suitable for additional helium IGM UV-spectroscopy studies. 
These are SDSS 0139-0847 ($z=3.13$), SDSS 0941+5607 ($z=3.79$), SDSS 
1006+3705 ($z=3.20$), SDSS 1007+4723 ($z=3.41$), SDSS 1009+3917 
($z=3.83$), SDSS 1137+6237 ($z=3.78$), SDSS 1442+0920 ($z=3.53$), SDSS 
2200+0008 ($z=3.13$), and SDSS 2251-0857 ($z=3.12$).  One case, SDSS 
0808+4550 (z=3.15) appears strongly absorbed with a sharp break 
$\sim65$\AA\ redward of \ion{He}{2}; perhaps this is due to a strong low-redshift 
hydrogen absorber at $z=0.45$. Somewhat more ambiguous are SDSS 0056-0941 
($z=3.25$) and our highest-redshift case, SDSS 1319+5202 ($z=3.90$), both 
of 
which appear to have more gradual absorption starting $\sim$60\AA\ redward 
of \ion{He}{2}. None of our spectra show highly significant flux 
blueward of the \ion{He}{2} break, consistent with the IGM helium having a 
non-negligible neutral or singly-ionized fraction in the broad redshift 
range probed. However, the spectra are of very low S/N, and in the 
lower-redshift cases there are very few data below the break.

Therefore, our approximate efficiency in identifying far-UV-bright quasars is of order 50\% (raw 
success rate among {\it HST}-observed cases) up 
to 90\% 
(using the criteria of the higher-confidence catalogs of section 2). Our 
efficiency in identifying new \ion{He}{2} quasar sightlines is $\sim 35\%$ 
(raw success rate among {\it HST}-observed cases) to $\sim65\%$ (using the 
higher-confidence criteria of the catalogs presented in section 2). These 
efficiencies are also consistent with our
spurious
superposition estimates of section 2.
But note that even among the higher 
confidence subset, reconnaissance {\it HST} observations to confirm flux all 
the way down to \ion{He}{2} are desired before undertaking lengthy follow-up 
UV spectral observations. In any case, our efficiency is confirmed to be 
an order of magnitude better than previous searches for \ion{He}{2} quasars, at
$\sim35-65\%$.

Some further caution is also appropriate in attributing any precise 
efficiency expectation to the other quasars cataloged in section 2, but 
not yet 
observed with {\it HST}. The quasars chosen 
for initial {\it HST} reconnaissance spectra are not a random sample for several 
reasons: (1) On average, the UV-brighter 
SDSS objects (as indicated by {\it GALEX}) were preferentially selected (see 
Figure 3) for 
spectroscopic observation with ACS, and thus constitute a potentially 
biased 
subset of our Table \ref{tab:TargetList} 
catalog. 
(2) Competing somewhat with the last bias is the 
need
to extend \ion{He}{2} studies to higher redshifts, especially to confirm, e.g.,
the
damped red 
profile expected at the reionization epoch; we thus often chose for {\it HST} 
prism reconnaissance the 
UV-brightest object in a given higher-redshift bin, in preference 
over 
adding many more UV-brightest cases at $z\sim3$, where there are several 
very well-studied quasar sightlines in the literature (see Figure 3, 
large green circles).
(3) We also excluded from {\it HST} observations quasars that were strongly 
absorbed 
in their SDSS optical spectra.
While 
our {\it HST} data thus demonstrate an unprecedentedly high efficiency for the
UV-brighter (and optically cleanest) 
objects, they do not directly confirm the fraction of 
fainter 
{\it GALEX} objects in our catalog that are true SDSS/{\it GALEX} matches, nor assess 
the
match efficiency of heavily (optically) absorbed cases such as BALQSOs.
That there are two quasars confirmed in our reconnaissance {\it HST} 
observations 
(SDSS 1010+3917 and 
SDSS 1137+6237) that appear in the GR1 catalog, but not in GR2/GR3 {\it GALEX} 
catalogs, suggests at least modest 
incompleteness as one reaches the {\it GALEX} UV depth limits (though 
completeness is not a consideration for our primary \ion{He}{2} science).

On the other hand, the UV-brightest of our new spectroscopically 
confirmed \ion{He}{2} quasars, such as SDSS 1006+3705 at
$z=3.20$, are comparable to the best of the venerable well-studied
\ion{He}{2} quasars. The far-UV (restframe) flux of this latter quasar is of 
order
$10^{-16}$~erg~s$^{-1}$~cm$^{-2}$~\AA$^{-1}$, similar to Q0302-003, for 
example.
There are a number of other such likely cases in our catalogs
presented in section 2, 
though many of the very UV-brightest cases are in the 
already reasonably well-studied 
$z<3.1$ regime (Figure 3); future studies may benefit from exploring 
the UV-brightest quasars in a given redshift regime.
But even the bulk of our cataloged likely SDSS/{\it GALEX} 
matches are 
well within expectations for future detailed observations with {\it HST}.
For example, the pre-flight COS estimated time
calculator indicates that 1-orbit observations of objects with UV fluxes
down to $\sim 5 \cdot 10^{-18}$~erg~s$^{-1}$~cm$^{-2}$~\AA$^{-1}$ would
yield spectra of sufficient quality to sample a proximity zone
near a \ion{He}{2}
Ly$\alpha$ break (S/N$\sim 5$ for 3 bins across a typical 20\AA\ wide,
$\tau \sim 1$ proximity zone, for objects in most of 
our redshift range).
The great bulk---about 97\%---of the objects in Tables 
\ref{tab:TargetList} and 2 satisfy such criteria. 

Comparison of Figures 1 and 3 also allows one to infer another result, 
perhaps not 
widely recognized. Most historical UV searches for \ion{He}{2} quasars emphasized 
searches mainly of the optically brightest quasars. While there is  
a correlation, these figures highlight that the optical is often not a 
good 
predictor of surviving quasar far-UV flux; many of the cases confirmed as 
far-UV 
bright in our {\it HST} reconnaissance are optically comparatively faint. 
The additional use of {\it GALEX} selection is 
essential to efficiently cull to the best new sightlines.

Our {\it HST} prism spectra are presented herein primarily to
verify the utility of our SDSS/{\it GALEX} selection approach (with initial 
efficiency estimates), and of course to definitively verify at least 9 new 
clean \ion{He}{2} quasar sightlines. However, the (admittedly modest S/N)
individual spectra displayed in Figure 5 also already suggest some 
diversity in 
far-UV 
(restframe) spectral character worthy of further study.
The 9 aforementioned quasars/sightlines appear
fairly clean and not too dissimilar to earlier \ion{He}{2} quasars. However,
there are three cases (SDSS 0056-0941, 0808+4550, 1319+5202) that are strongly
absorbed starting $\sim$60\AA\ redward of the \ion{He}{2} break. [It may 
be worth
recalling that other optically bright quasars, such as UM670 and OQ172,
considered but rejected in early {\it HST} \ion{He}{2} searches, 
also appear to have
redward breaks \citep{L94,L95,jak93}]. And in our own {\it HST} confirmation 
there are at least
two quasars (SDSS 1006+3705, 1442+0920) perhaps with unusually strong
far-UV emission features. This diversity reinforces the utility of initial
reconnaissance {\it HST} observations in planning for more detailed UV 
spectral
follow-on observations. It also emphasizes the necessity of obtaining a 
large sample of
independent \ion{He}{2} quasar sightlines, spanning a broad redshift range, to
allow adequate averaging over cosmic variance and/or individual-object
pathology in helium studies. (Of course even if some or all of the
redward-absorbed cases discussed herein are ultimately understood as
unsuitable for IGM \ion{He}{2} studies, they may still be potentially
interesting for studies of low-redshift hydrogen absorbers, for example).

It may also be of interest to broadly compare our ensemble prism spectra
here with the ACS prism spectrum (same instrumental setup, but much longer 
exposure and better S/N) of the intriguing \ion{He}{2} quasar SDSS J1711+6052 
detailed in \citet{Z08}.
In that paper, we discuss the possibility that the spectral character
of SDSS 1711+6052
observed redward of the expected (1465 \AA) \ion{He}{2} break location
might be associated with the red 
wing of \ion{He}{2} Ly$\alpha$ emission, 
strongly absorbed by a very broad damped profile.
That damped profile might arise 
from low-redshift hydrogen, or interestingly perhaps from a surprisingly 
dense (column density $\sim 10^{22}$~cm$^{-2}$) helium 
Ly$\alpha$
absorption at high redshift, for example.
An ensemble stack of the UV spectrum of seven of our new clear quasar 
sightlines at $z<3.5$ (in the observed wavelength frame) is shown in Figure 6.
This stacked ensemble spectrum is largely devoid of very strong features 
in the 
instrumental frame near 1400-1500\AA, i.e., devoid of very strong 
features 
near the observed wavelength of the possibly damped profile seen in SDSS 
1711+6052. This stacked ensemble ACS spectrum then would seem to at least 
argue 
strongly that the possibly damped profile  
absorption feature discussed in 
\citet{Z08} for SDSS J1711+6052 is unlikely merely some instrumental 
artifact of ACS prisms.

\section{Summary and Concluding Remarks}

We provide a catalog of 200 higher-confidence ($\sim$70\% secure) SDSS 
quasars at $z>2.78$ that are likely statistically associated with 
GALEX ultraviolet sources. We provide an additional set of 112 similar quasars, also $z>2.78$ but on average lower redshift, from consideration of 
the  
\citet{ver06} quasar compilation. Our catalogs herein comprise very large 
resource 
lists (two orders of magnitude larger than previous similarly dedicated 
efforts) of 
likely 
far-UV (restframe) bright quasars, potentially of interest for a range of 
studies from far-UV SED, emission line, and BALQSO spectral studies, to 
background probes
of intervening clouds including \ion{H}{1} at low-redshift, to our own primary 
interest in
the \ion{He}{2} Gunn-Peterson test.

We report reconnaissance UV spectral observations from {\it HST} of 24 
SDSS/{\it GALEX} selected quasars, chosen to encompass a wide redshift range,
finding that half are indeed far-UV bright. At 
least 9
new quasars are confirmed in our {\it HST} prism spectra as suitable new 
sightlines
for \ion{He}{2} Gunn-Peterson studies. The {\it HST} data also allow us to confirm the 
even 
higher efficiency of our refined selection approach, which generated the catalogs of section 2.
In either case, our combined 
SDSS/{\it GALEX}/{\it HST}-reconnaissance selection efficiency is 
an order of magnitude better than previous similar \ion{He}{2} quasar searches, 
and we herein more than double the number of spectroscopically confirmed 
\ion{He}{2} quasars.

Our {\it HST} prism spectra show some diversity, reaffirming
the need for UV 
reconnaissance spectral observations to confirm 
that UV flux extends
all the way down to \ion{He}{2} (even in cases detected in broadband from {\it GALEX}). 
The spectral diversity found also confirms the
necessity of identifying multiple independent clean sightlines, covering 
a broad redshift range, to limit cosmological variance and 
individual-object pathology in seeking global understanding from helium 
IGM 
studies.
An observed-frame ensemble stack of our {\it HST} reconnaissance spectra 
suggests that the  
strong absorption seen in SDSS1711+6052 by \citet{Z08} redward of the 
\ion{He}{2} break is unlikely to be an instrumental artifact.

To probe IGM \ion{He}{2} in detail, including the epoch of \ion{He}{2}
reionization, requires a large enough sample of clean quasar sightlines, 
extending 
over a broad 
redshift range, in order to be limited by cosmic
variance rather than individual peculiarities or a clumpy IGM. With 
such a large, broad-redshift sample, concerns of 
a possible systematic bias due to clear sightline selection
may be addressable using the observed sightlines to calibrate
large-scale numerical simulations of reionization \citep[e.g.,][]{W08}.
The range of redshift encompassed by our sample may also allow future
observations to constrain percolation models of reionization, such as
the one discussed by \citet{P07}.

The empirical surprises at high redshift \citep{Z08}, theoretical 
uncertainties in interpreting the past low redshift results \citep{mei08},
and the broad goal of identifying the epoch of helium 
reionization through the anticipated redward damped \ion{He}{2} profile 
signature, all argue strongly for the need to emphasize \ion{He}{2} Gunn-Peterson 
studies beyond the $z\sim3$ regime of the bulk of the venerable past 
studies. Our catalogs of hundreds of far-UV-bright quasars, and even more 
directly our {\it HST} spectroscopic confirmations of at least 9 new \ion{He}{2} 
quasars (along 
with several more from our earlier similar work), 
provide basic resource lists extending to such higher 
redshifts for future detailed (good S/N and spectral resolution) UV 
spectral science investigations of helium in the IGM.

\acknowledgments

Funding for the SDSS and SDSS-II has been provided by the Alfred P. Sloan 
Foundation, the Participating Institutions, the National Science 
Foundation, the U.S. Department of Energy, the National Aeronautics and 
Space Administration, the Japanese Monbukagakusho, the Max Planck 
Society, and the Higher Education Funding Council for England. The SDSS 
Web Site is http://www.sdss.org/.

The SDSS is managed by the Astrophysical Research Consortium for the 
Participating Institutions. The Participating Institutions are the 
American Museum of Natural History, Astrophysical Institute Potsdam, 
University of Basel, University of Cambridge, Case Western Reserve 
University, University of Chicago, Drexel University, Fermilab, the 
Institute for Advanced Study, the Japan Participation Group, Johns 
Hopkins University, the Joint Institute for Nuclear Astrophysics, the 
Kavli Institute for Particle Astrophysics and Cosmology, the Korean 
Scientist Group, the Chinese Academy of Sciences (LAMOST), Los Alamos 
National Laboratory, the Max-Planck-Institute for Astronomy (MPIA), the 
Max-Planck-Institute for Astrophysics (MPA), New Mexico State University, 
Ohio State University, University of Pittsburgh, University of 
Portsmouth, Princeton University, the United States Naval Observatory, 
and the University of Washington.

We gratefully acknowledge support from NASA/{\it GALEX} Guest Investigator 
grants NNG05GE12G and NNG06GD03G.
Support for {\it HST} Program numbers 10132, 10907, and 11215 was 
provided by NASA
through grants from the Space Telescope Science Institute, which is
operated by the Association of Universities for Research in Astronomy,
Incorporated, under NASA contract NAS5-26555.

\clearpage


\begin{deluxetable}{lrrcccccc}
\tabletypesize{\footnotesize}
\rotate
\tablewidth{550pt}

\tablecaption{Catalog of Candidate Far-UV-Bright SDSS/{\it GALEX} Quasars 
\label{tab:TargetList}}

\tablehead{\colhead{Name} & \colhead{RA} & \colhead{Dec} & 
\colhead{Redshift} & \colhead{z-band} & \colhead{NUV flux} & \colhead{FUV 
flux} & \colhead{Insp.\tablenotemark{b}} & \colhead{Obs.\tablenotemark{c}} \\ 
\colhead{} & \colhead{(J2000)} & \colhead{(J2000)} & \colhead{} & \colhead{(mag)} & \colhead{($10^{-17}$)\tablenotemark{a}} & \colhead{($10^{-17}$)\tablenotemark{a}} & \colhead{} & \colhead{} } 
\startdata
SDSSJ000303.35-105150.7 & 0.763963 & -10.864074 & 3.65 & 19.00 & 0.803 & 3.209 & 6 & 0 \\ 
SDSSJ000316.39-000732.4 & 0.818289 & -0.125675 & 3.17 & 19.99 & 3.218 & ... & 0 & 0 \\ 
SDSSJ001632.44+001421.6 & 4.135159 & 0.239336 & 3.33 & 20.04 & 2.449 & ... & 0 & 0 \\ 
SDSSJ001641.17+010045.3 & 4.171557 & 1.012579 & 3.05 & 19.28 & ... & 5.232 & 1 & 0 \\ 
SDSSJ003420.62-010917.3 & 8.585921 & -1.154819 & 2.85 & 20.05 & 1.903 & 1.849 & 0 & 0 \\ 
SDSSJ003930.32+001754.2 & 9.876321 & 0.298402 & 2.90 & 20.00 & 0.534 & ... & 0 & 0 \\ 
SDSSJ004323.43-001552.6 & 10.847631 & -0.264620 & 2.80 & 18.14 & 7.721 & ... & 8 & 0 \\ 
SDSSJ005401.48+002847.8 & 13.506173 & 0.479941 & 3.41 & 19.80 & 4.985 & ... & 3 & 0 \\ 
SDSSJ005653.26-094121.9 & 14.221906 & -9.689411 & 3.25 & 19.77 & 3.144 & 3.103 & 0 & 1 \\ 
\enddata

\tablecomments{Excerpt of table. Full version available online in machine-readable format.}

\tablenotetext{a}{erg s$^{-1}$ cm$^{-2}$ \AA$^{-1}$, from 
{\textit{GALEX}}}
\tablenotetext{b}{The numeric inspection flag denotes possible problems with the 
object. The individual values are: $1$ for a very near neighbor in the SDSS 
image, $2$ for a probable LLS or DLA in the SDSS spectrum, $4$ for a BAL, and $8$ for a possible BAL. These individual flags are additive for a given object (e.g., 6 denotes a BAL with an LLS or DLA).}
\tablenotetext{c}{The observation flag is 0 if the object has not been observed with {\it HST}, 1 if it 
has 
been observed with FUV (to near 304\AA\ rest) spectroscopy through {\it HST} cycle 16, and 2 if 
it 
has been observed less conclusively, i.e., NUV spectroscopy, UV imaging, or pre-COSTAR observation.}

\end{deluxetable}

\begin{deluxetable}{lrrlcccc}
\tabletypesize{\footnotesize}
\rotate
\tablewidth{550pt}
\tablecaption{Catalog of Candidate Far-UV-Bright (non-SDSS) 
VCV/{\it GALEX} Quasars
\label{tab:TargetListVeron}}

\tablehead{\colhead{Name} & \colhead{RA} & \colhead{Dec} & \colhead{Redshift} & \colhead{V\tablenotemark{a}} & \colhead{NUV flux} & \colhead{FUV flux}  & \colhead{Obs.\tablenotemark{c}} \\ 
\colhead{} & \colhead{(J2000)} & \colhead{(J2000)} & \colhead{} & \colhead{(mag)} & \colhead{($10^{-17}$)\tablenotemark{b}} & \colhead{($10^{-17}$)\tablenotemark{b}} & \colhead{} } 

\startdata
TEX0004+139 & 1.7396 & 14.2628 & 3.2 & 19.90 & 4.73 & ... & 0 \\
CXOMPJ00184+1629 & 4.6125 & 16.4831 & 2.828 & R21.27 & 0.37 & ... & 0 \\
CXOMPJ00189+1629 & 4.7288 & 16.4978 & 2.95 & R21.78 & 0.88 & ... & 0 \\
Q0016-3936 & 4.7992 & -39.3278 & 3. & *19.79 & 25.05 & 38.29 & 0 \\
2QZJ002416-3149 & 6.0675 & -31.8289 & 2.846 & *20.24 & 4.44 & ... & 0 \\
Q0023-4013 & 6.5912 & -39.9514 & 3. & *19.64 & 3.86 & ... & 0 \\
Q0026-3934 & 7.1929 & -39.3050 & 2.91 & *19.59 & 1.64 & ... & 0 \\
Q0027-4132 & 7.4904 & -41.2589 & 2.79 & *19.70 & 3.80 & ... & 0 \\
2QZJ003130-3033 & 7.8750 & -30.5558 & 2.811 & *19.16 & 2.13 & ... & 0 \\
2QZJ003447-3048 & 8.6967 & -30.8033 & 2.785 & *19.97 & 3.99 & ... & 0 \\
\enddata

\tablecomments{Excerpt of table. Full version available online in machine-readable format.}

\tablenotetext{a}{V magnitude unless otherwise marked as: 
photographic (*), blue (B), red (R), infrared (I J K), or photographic O- or J- plates (O).}
\tablenotetext{b}{erg s$^{-1}$ cm$^{-2}$ \AA$^{-1}$, from {\textit{GALEX}}}
\tablenotetext{c}{The observation flag is 0 if the object has not been observed with {\it HST}, 1 if it 
has 
been observed with FUV (to near 304\AA\ rest) spectroscopy through {\it HST} cycle 16, and 2 if 
it 
has been observed less conclusively, i.e., NUV spectroscopy, UV imaging, or pre-COSTAR observation.}

\end{deluxetable}

\begin{deluxetable}{lrrcccclc}
\tabletypesize{\scriptsize}
\rotate
\tablewidth{600pt}
\tablecaption{Quasars/Sightlines with Confirming ACS Reconnaissance 
UV Spectra\label{tab:ACSObs}}

\tablehead{\colhead{Name} & \colhead{RA} & \colhead{Dec} & \colhead{Redshift} & \colhead{NUV flux} & \colhead{FUV flux} & \colhead{Exp. Time\tablenotemark{b}} & \colhead{Obs. Date} & \colhead{He II Ly$\alpha$ break} \\
\colhead{} & \colhead{(J2000)} & \colhead{(J2000)} & \colhead{} & \colhead{($10^{-17}$)\tablenotemark{a}} & \colhead{($10^{-17}$)\tablenotemark{a}} & \colhead{(s)} & \colhead{} & \colhead{} }

\startdata
SDSSJ005653.26-094121.9 & 14.221906 & -9.689411 & 3.25 & $3.15 \pm 0.70$ & $3.10 \pm 0.91$ & 3200 & 2007 Jul 18 & ? \\
SDSSJ013900.79-084720.6 & 24.753291 & -8.789060 & 3.13 & ... & $4.92 \pm 0.95$ & 4240 & 2007 Jul 19 & yes \\
SDSSJ080856.19+455006.8 & 122.234134 & 45.835213 & 3.15 & $0.58 \pm 0.25$ & $4.4 \pm 1.3$ & 4350 & 2007 Dec 3 & no \\
SDSSJ094102.52+560706.6 & 145.260480 & 56.118488 & 3.79 & $1.34 \pm 0.82$ & ... & 3630 & 2007 Apr 15 & yes \\
SDSSJ100610.55+370513.9 & 151.543968 & 37.087186 & 3.20 & $16.3 \pm 2.4$ & $6.1 \pm 2.9$ & 4300 & 2007 Nov 10 & yes \\
SDSSJ100745.90+472321.1 & 151.941266 & 47.389197 & 3.41 & $2.6 \pm 1.1$ & $8.2 \pm 3.4$ & 4350 & 2007 Dec 19 & yes \\
SDSSJ100956.05+391718.4 & 152.483550 & 39.288447 & 3.83 & $5.3 \pm 1.9$\tablenotemark{c} & ... & 3200 & 2007 Nov 29 & yes \\
SDSSJ113721.72+623707.2 & 174.340509 & 62.618667 & 3.78 & $4.5 \pm 2.6$\tablenotemark{c} & ... & 3200 & 2007 Oct 29 & yes \\
SDSSJ131914.20+520200.1 & 199.809186 & 52.033356 & 3.90 & ... & $10.4 \pm 4.2$ & 3200 & 2007 Nov 9 & ? \\
SDSSJ144250.12+092001.5 & 220.708836 & 9.333762 & 3.53 & $5.4 \pm 1.2$ & ... & 3200 & 2007 Mar 12 & yes \\
SDSSJ220040.92+000832.2 & 330.170480 & 0.142289 & 3.13 & $6.31 \pm 0.53$ & $11.0 \pm 1.2$ & 4240 & 2008 Jun 19 & yes \\
SDSSJ225117.80-085722.8 & 342.824147 & -8.956327 & 3.12 & $5.52 \pm 0.39$ & $11.5 \pm 1.6$ & 4280 & 2007 May 13 & yes \\
\enddata

\tablenotetext{a}{erg s$^{-1}$ cm$^{-2}$ \AA$^{-1}$, from {\it GALEX}}
\tablenotetext{b}{Some quasars had one or two particularly noisy or misaligned sub-exposures. Our analysis does not use these sub-exposures, nor have they been included in these tabulated exposure times.}
\tablenotetext{c}{Flux from {\it GALEX} GR1, since this source is not in GR2/GR3.}

\end{deluxetable}

\clearpage

\begin{figure}
\figurenum{1}
\epsscale{0.8}
\plotone{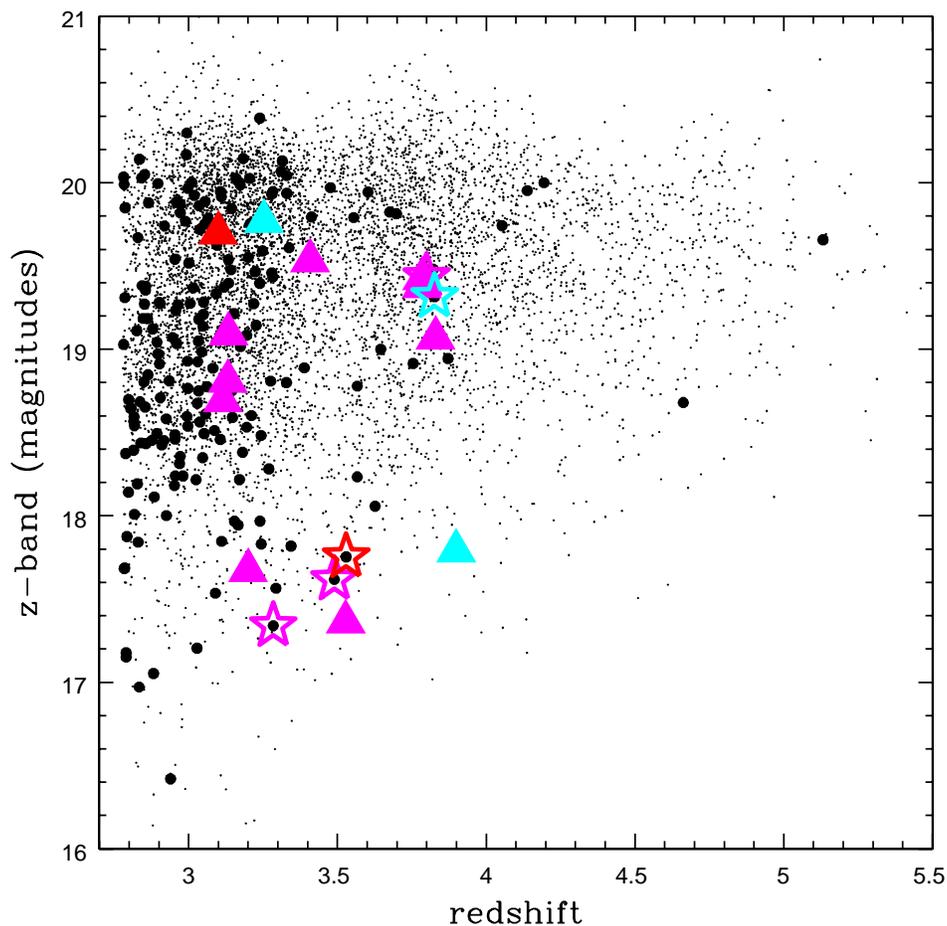}
\caption{
A portion of the redshift-magnitude diagram for SDSS quasars (smaller 
black points). There are many SDSS quasars of potential interest for 
helium absorption studies, e.g., with 7800 considered herein at $z>2.78$. 
The two hundred SDSS/{\it GALEX} matches cataloged here in section 2 as 
higher-confidence far-UV-bright quasars are highlighted as medium, filled 
black circles. (The symbols in color denote quasars having UV 
verification spectra, e.g., in anticipation of comparisons with Figure 3 
and related discussion in the later sections of the text; the 
color-symbol coding here is the same as detailed in the Figure 3
caption).}
\end{figure}

\begin{figure}
\figurenum{2}
\epsscale{0.9}
\plotone{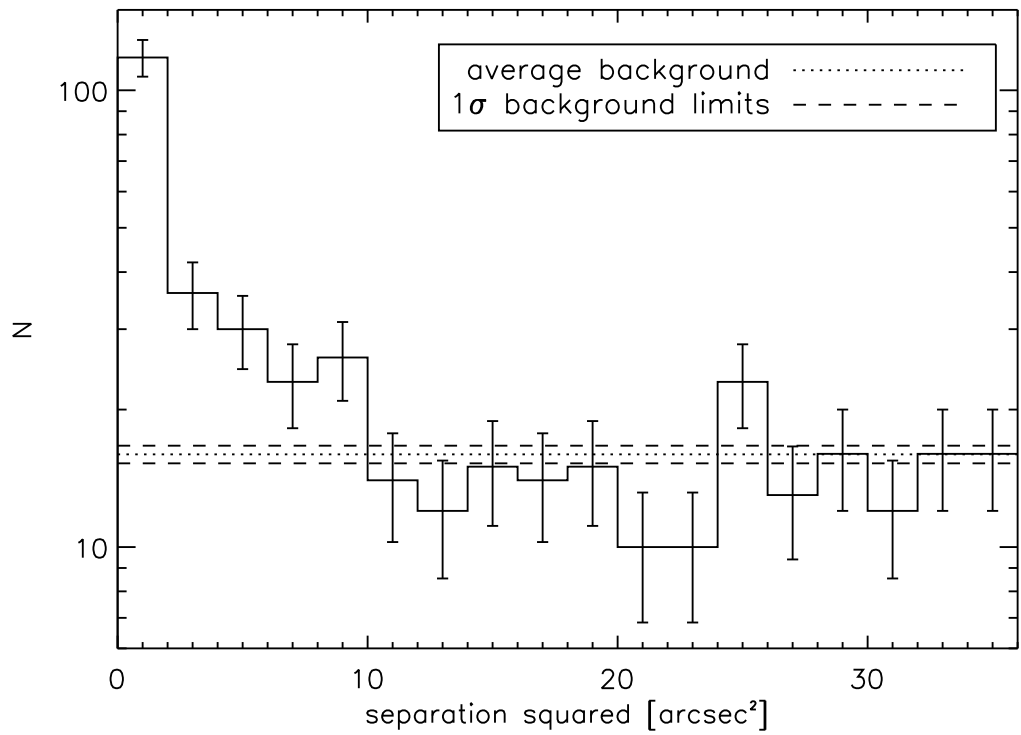}
\caption{SDSS-{\it GALEX} source separation, with the background 
spurious match rate shown. The histogram shows the 419 $r \leq 6''$ 
matches, while the error bars are simply $\sqrt{N}$ Poisson noise 
for each bin.
\label{fig:SeparationHist}}
\end{figure}

\begin{figure}
\figurenum{3}
\epsscale{0.8}
\plotone{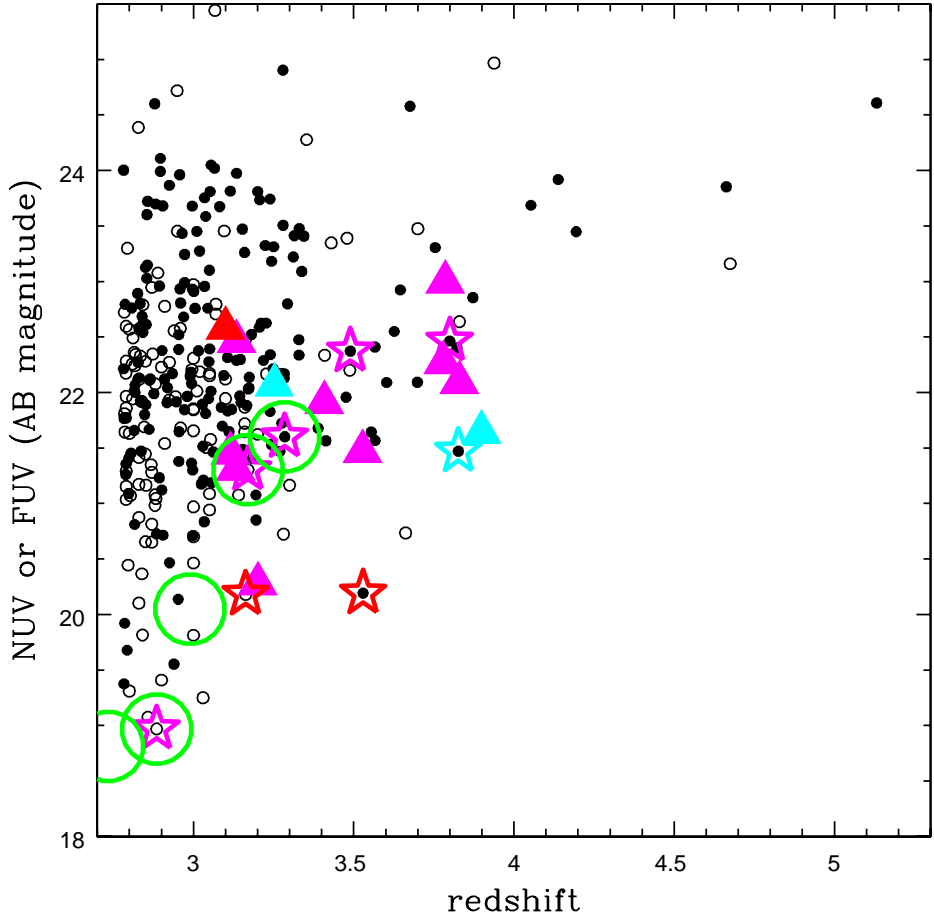}
\caption{
Redshift/UV-magnitude diagram for our 200 higher-confidence 
far-UV-bright quasars at $z>2.78$ from SDSS (medium, filled black circles),
and the 112 similar cases derived from VCV \citep{ver06} quasars (medium,
open black circles). Where there is
a UV detection in only one of the two {\it GALEX} passbands, we plot the 
relevant associated AB magnitude, NUV or FUV 
as appropriate; for cases detected in both FUV and 
NUV bandpasses we plot the {\it GALEX} AB magnitude of the brighter of the two. The 
large triangles depict the 12 {\it HST}/prism-observed SDSS/{\it GALEX} quasars 
thus far confirmed as bright in the far-UV restframe (including two found 
only in GR1);
magenta triangles depict the 9 whose {\it HST} prism spectra 
confirm them as excellent new clean sightlines, cyan triangles are 2 cases 
which may be absorbed just above the \ion{He}{2} break and require further 
UV spectroscopic scrutiny, and the red triangle shows one case where 
our prism spectra already convincingly show strong and sharp absorption 
significantly redward of the \ion{He}{2} break. For comparison, the open 
starred symbols show the data for the previously published confirmed or 
strongly suspected \ion{He}{2} quasars; the color coding is the same as 
for the triangles. Large green circles denote the 5 \ion{He}{2} quasars with 
extant very high-quality (in both resolution and S/N) UV spectra.} 
\end{figure}

\begin{figure}
\figurenum{4}
\epsscale{0.9}
\plotone{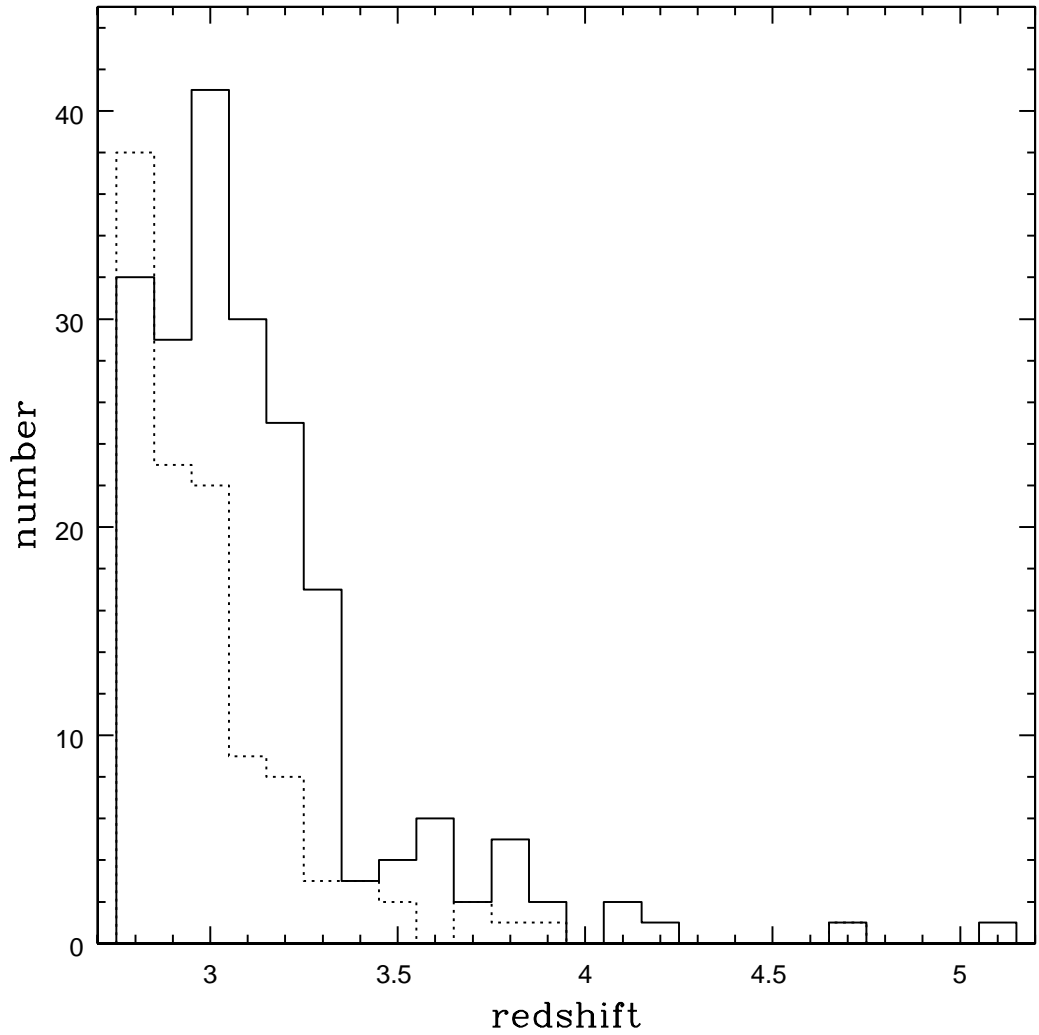}
\caption{Redshift distributions of likely far-UV-bright quasars
from {\it GALEX} GR2/GR3 and SDSS DR5/DR6 (solid line), or VCV
(dotted line) 
quasars \citep{ver06} at $z>2.78$.}
\end{figure}

\begin{figure}
\figurenum{5a}
\epsscale{1.0}
\plotone{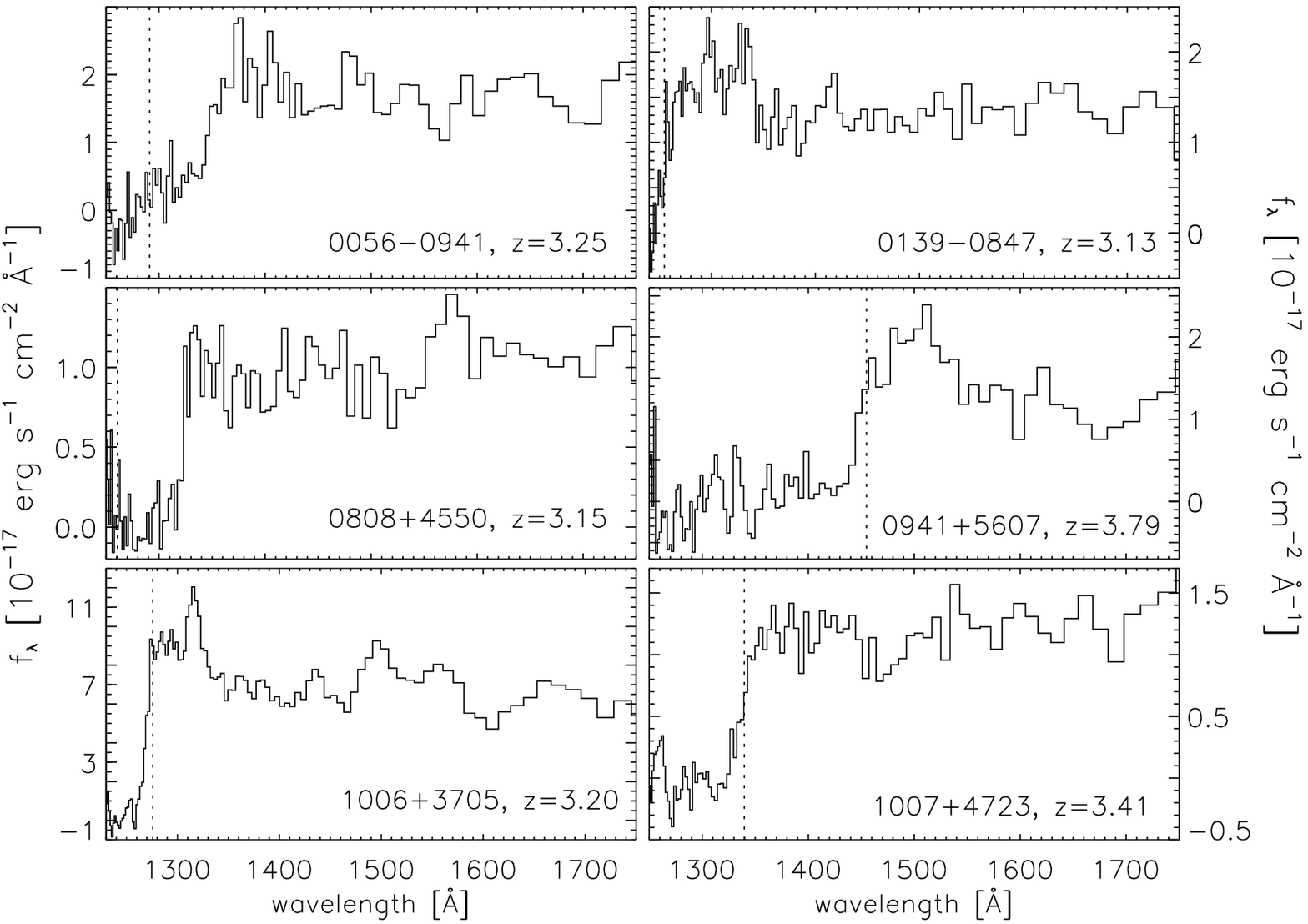}
\caption{ACS spectra for the first 6 objects in Table \ref{tab:ACSObs}.
The vertical dotted line shows the anticipated position of \ion{He}{2} Ly$\alpha$ for
the quasar redshift. Instrumental resolution is about 2 bins.
\label{fig:Spec1-6}}
\end{figure}

\begin{figure}
\figurenum{5b}
\epsscale{1.0}
\plotone{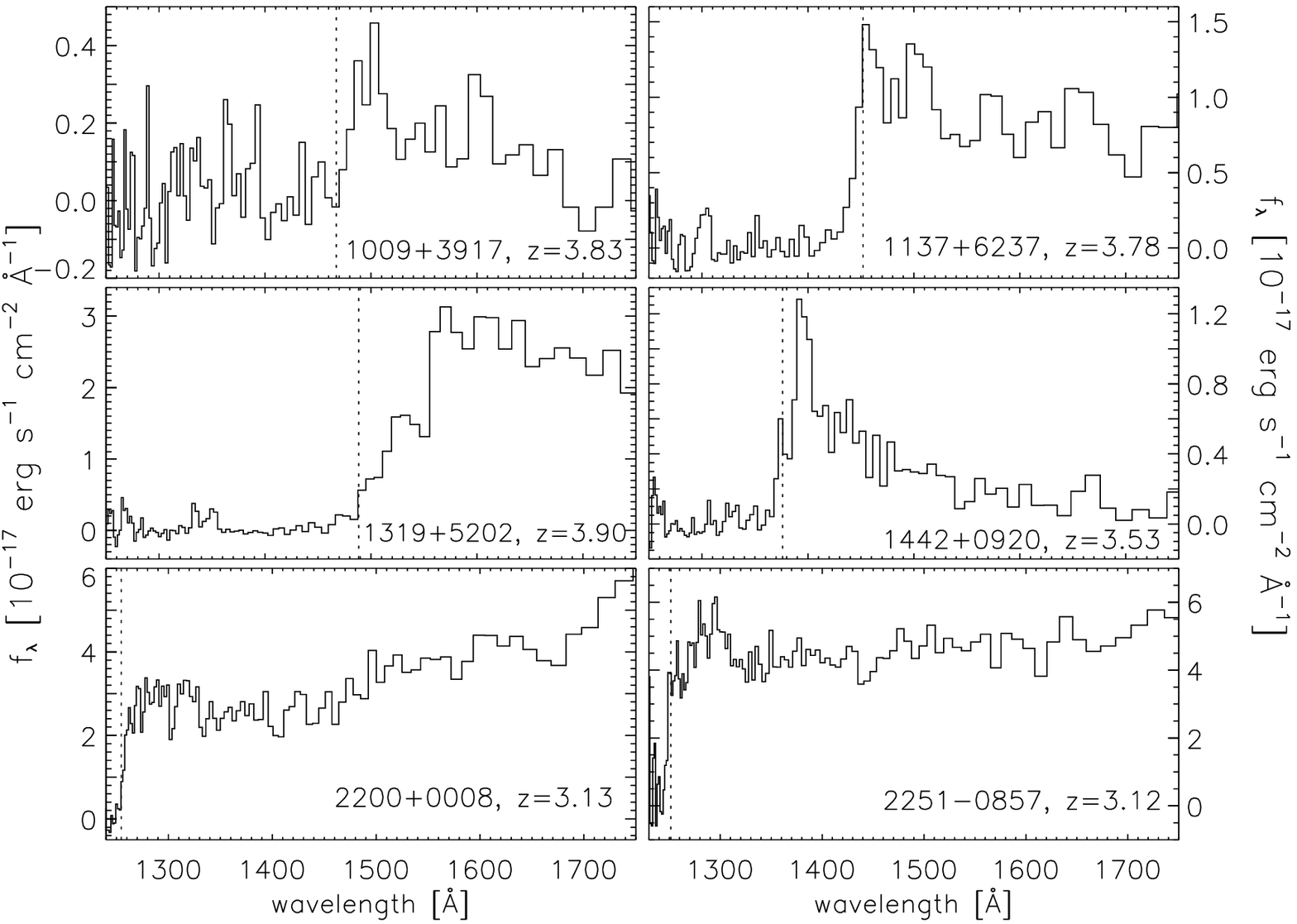}
\caption{ACS spectra for the last 6 objects in Table \ref{tab:ACSObs}. The
vertical dotted line shows the anticipated position of \ion{He}{2} Ly$\alpha$ for the
quasar redshift. Instrumental resolution is about 2 bins.
\label{fig:Spec7-11}}
\end{figure}

\begin{figure}
\figurenum{6}
\epsscale{0.8}
\plotone{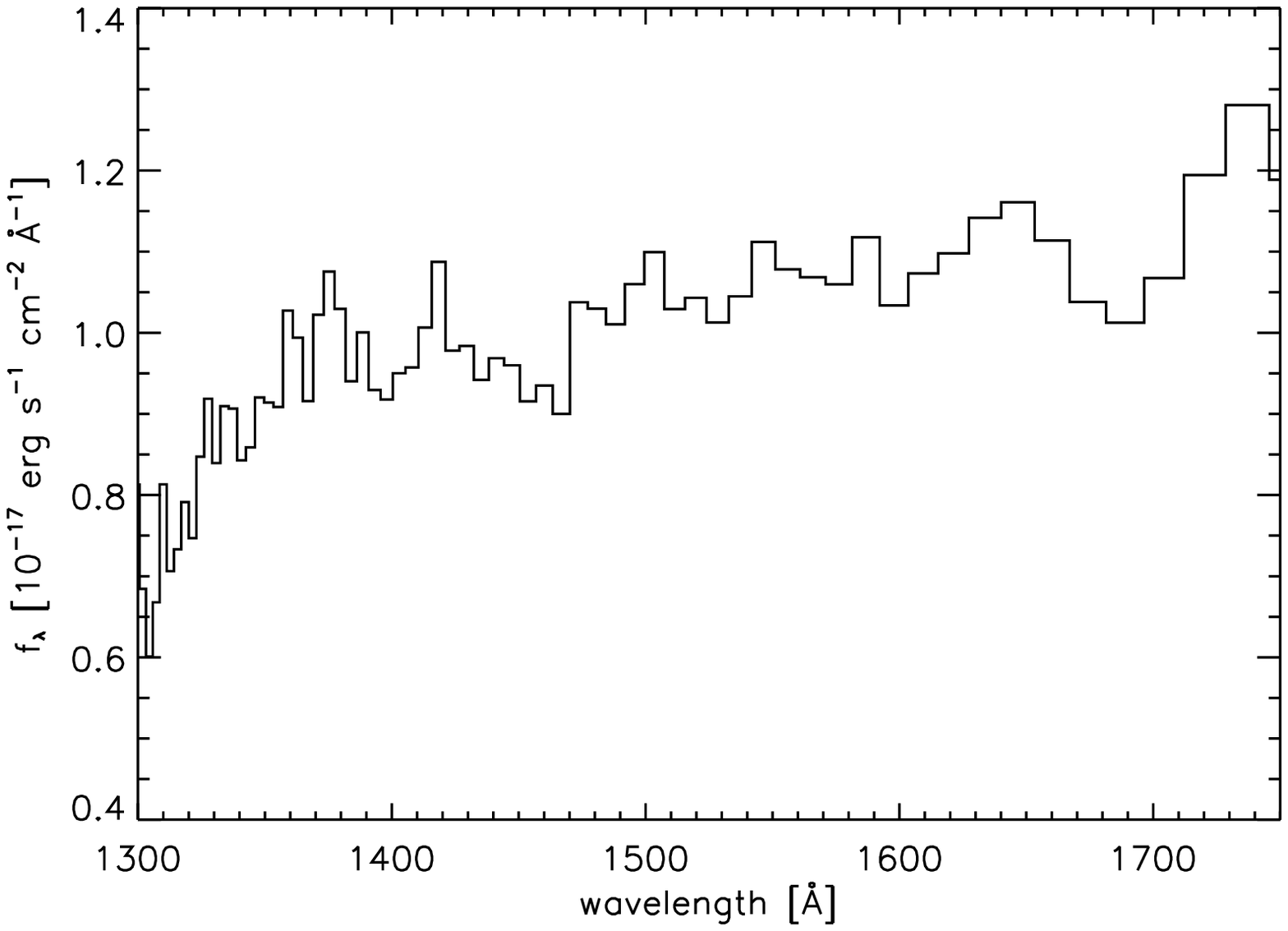}
\caption{Seven of our UV prism spectra coadded/stacked in the instrument 
(observed) frame,
weighted by flux. To obtain coverage of much of the wavelength range, the
included objects are limited to those at lower redshift, 
$z<3.5$ (\ion{He}{2} Ly$\alpha$ at $\lesssim$1350\AA). The slightly red slope is
largely attributable to a single object, 2200+0008.
\label{fig:Stack7Inst}}
\end{figure}

\end{document}